# Ray computational ghost imaging based on rotational modulation method

Zhi Zhou,[1] Sangang Li,[1,2,*] Shan Liao,[1] Sirun Gong,[1] Rongrong Su,[1] Chuxiang Zhao,[1] Li Yang,[3] Qi Liu,[4] Yucheng Yan,[5] Mingzhe Liu[1] and Yi Cheng[1]

*1 College of Nuclear Technology and Automation Engineering, Chengdu University of Technology, Chengdu, 610059, China*
*2 Applied Nuclear Technology in Geosciences Key Laboratory of Sichuan Province, Chengdu University of Technology, Chengdu, 610059, China*
*3 Institute of Plasma Physics, Hefei Institutes of Physical Science, Chinese Academy of Science, Hefei 230031, China*
*4 Sichuan University of Science and Engineering, Zigong, 643000 China*
*5 The Engineering & Technical College of Chengdu University of Technology, Leshan 614000, China*
*\* lisangang@cdut.edu.cn*

**Abstract:** The CGI (CGI) has the potential of low cost, low dose, and high resolution, which is very attractive for the development of radiation imaging field. However, many sub-coding plates must be used in the modulation process, which greatly affects the development of CGI technology. In order to reduce the coding plates, we refer to the rotation method of computed tomography (CT), then propose a novel CGI method based on rotational modulation method of a single-column striped coding plate. This method utilizes the spatial variation of a single sub-coding plate (rotation) to realize multiple modulation of the ray field and improves the utilization rate of a single sub-coding plate. However, for this rotation scheme of CGI, the traditional binary modulation matrix is no longer applicable. To obtain the system matrix of the rotated striped coding plate, an area model based on beam boundaries is established. Subsequently, numerical and Monte Carlo simulations were conducted. The results reveal that our scheme enables high-quality imaging of N×N resolution objects using only N sub-coding plates, under both full-sampling and under-sampling scenarios. Moreover, our scheme demonstrates superiority over the Hadamard scheme in both imaging quality and the number of required sub-coding plates, whether in scenarios of full-sampling or under-sampling. Finally, an α ray imaging platform was established to further demonstrate the feasibility of the rotational modulation method. By employing our scheme, a mere 8 sub-coding plates were employed to achieve CGI of the radiation source intensity distribution, achieving a resolution of 8×8. Therefore, the novel ray CGI based on rotational modulation method can achieve high-quality imaging effect with fewer sub-coding plates, which has important practical value and research significance for promoting single-pixel radiation imaging technology.

## 1.Introduction

Computational ghost imaging (CGI), also known as single pixel imaging, has received extensive attention in recent years as a new imaging technology [1-8]. The essence of this technology is that the spatial information of source intensity is encoded by a modulator, then the intensity is collected after passing the object through a single pixel detector (bucket detector), and finally the intensity information and modulated spatial information are decoded to recover the object image.

It is precisely the "encoding-decoding" calculation form that allows a bucket detector rather than detector arrays to obtain the object images. So, the CGI can greatly reduce the requirements of detectors in imaging system.

In the field of ray CGI [9-17], these modulators (spatial light modulator (SLM) and digital micromirror (DMD)) generally used in light field cannot modulate radioactive ray particles with a certain penetration ability. The only way is to take advantage of the attenuation of rays. And a coding plate composed of many sub-coding plates with certain blocking ability is employed as a modulator [18]. However, in CGI, many numbers of sub-coding plates are required. This will greatly affect the cost and convenience of the radiation imaging system and transfer the saved cost from the manufacturing of the detector to that of the coding plate. Especially for the ray with strong penetration capability, the required coding plate has defects such as large thickness and many pixels, etc. These will result in high manufacturing cost and great manufacturing difficulty. In this case, how to reduce the number of pixels of the coding plate, to reduce the cost and difficulty, is an urgent problem to be solved. This is important significance for implementing ray CGI.

For this problem, scholars have carried out some research and got many achievements. Specifically, it can be divided into two ideas: optimizations of modulation scheme and measurement form.

Class I: optimization of modulation scheme [19-22].

From the perspective of information theory, the CGI can be regarded as the modulation and demodulation process of the target image information from the physical level. Under the same reconstruction algorithm, the modulation scheme, a coding plate, greatly affects the quality of a reconstructed image [19]. Among the coding plates, such as random and Hadamard coding plates, etc. Hadamard coding plate has perfect orthogonality which could avoid the correlation noise between each pixel. Especially, when it is combined with a compressed sensing technology or depth learning technology, small number of "high-quality" Hadamard sub-coding plates to reconstruct high-quality image can be used. The requirement of CGI system on the number of sub-coding plates can be reduced. For example, in 2017, the Russian Doll (RD) sampling sequence was designed by Sun according to structural characteristics of Hadamard matrix and total number of sub-coding plate blocks [20]. In his study, only 16% sub-coding plates can get a good reconstructed image. Furthermore, in 2019, YU proposed the "Cake Cutting" sequence that is sorted in ascending order according to the number of blocks of Hadamard basis, and high-quality imaging can be reconstructed at a smaller sampling rate [21,22].

On average, various under-sampling schemes mentioned above can reduce the number of measurements by about 85%. This could be accepted for a single pixel image with DMD or SLM modulation in visible light band, but the number of sub-coding plates are still too many for ray CGI. It is very difficult to manufacture so many sub-coding plates. And so many times of up and down translation seriously affect the measurement time. These severely restricts the development of single pixel radiation imaging technology.

Class II: optimization of measurement form [23-26].

In traditional CGI, a sub-coding plate will be discarded after each measurement. Then a new sub-coding plate for the next measurement will be moved to the same measurement position. But these sub-coding plates may have a characteristic of the same areas[9]. The characteristics can be used to reduce the total pixels of the coding plate. Representative schemes are following: In 2020, a rapid measurement scheme was proposed by Jiang [25]. A strip shaped random coding plate was used, and a column of the coding plate was moved in every measurement to realize the change of modulation effect. Only a coding plate of $N^2 \times (1+10\%)$ ($N^2$ represents the pixel number of a sub-coding plate) pixels combined with the neural network recovery algorithm were used to realize a reconstruction of targets. In addition, the measurement form that only one column is moved also occurred on the orthogonality coding plate. In 2021, the cyclic S-matrix was employed to form a new mask by Hdamovich [26]. The scheme only moving a column in each measurement can also realize the undistorted reconstruction of the target under full sampling. These schemes were originally designed for realizing a rapid measurement. They also directly reduced the required pixels of a coding plate and simplified the coding plate in ray CGI system. But, this switching scheme, moving one column, causes little change in the modulation effect, leading to a lot of degradation in the image quality, which limits further reduction of measurement times.

In conclusion, optimization of measurement form is an effective way to achieve lightweight and miniaturization of modulator components and a rapid measurement, but only horizontal moving measurement cannot guarantee the modulation effect. Here, this paper intends to improve measurement form through rotation measurements replacing a big part of horizontal measurements, and thus achieve equivalent modulation by spatial variation of a few sub-coding plates. The beam model was subsequently established to accurately compute the ray field after modulation by a rotating coding plate, thereby deriving the system matrix under this unique modulation. Ultimately, the feasibility of the proposed approach is validated through numerical simulations, Monte Carlo simulations, and practical experimentation.

## 2.Methods

### 2.1 Traditional ray CGI

The schematic of the traditional ray CGI device is shown in Fig. 1.

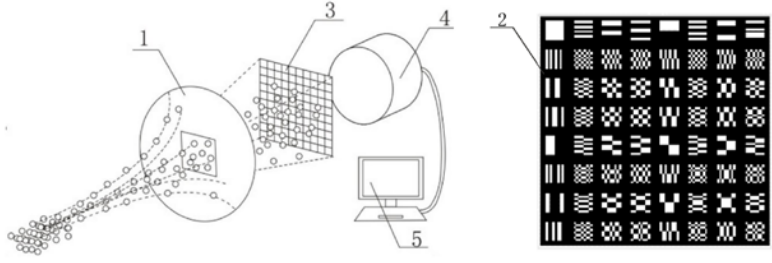

Fig. 1 Conceptual drawing of the experimental setup used for traditional ray CGI. [9] The left and right parts of the image show the overall layout and a partial enlargement of the equipment, respectively. The collimator (1) ensures that the particles produced by the ray source are only incident on the imaging area; The Hadamard coding plate (2) is composed of all the sub-coding plates, whose pattern is encoded on metal; The particles are modulated on the sub-coding plate (3), and the object is placed behind the sub-coding plate; the signal-pixel detector (4) collects the intensity of rays; a computer (5) is used to reconstruct the images.

To achieve CGI, during the intensity measurement phase, it is necessary to employ correlation modulation devices (such as DMD) to multiply alter the incident light field onto the object. This process yields multiple intensity measurement values. However, radiation particles possess robust penetrating capabilities. To alter the spatial distribution of the radiation field, different patterns in a metal mask (coding plate) are employed in different measurements. The part of the mask used in each measurement was called a sub-coding plate. During different measurements, the switching of sub-coding plates occurs through the horizontal/vertical movement of the coding plates.

In the right part of Fig.1, it is a coding plate designed to achieve an 8x8 resolution imaging, incorporating 64 sub-coding plates required for 64 measurements. After each measurement, it is necessary to shift the coding plate, causing different sub-coding plates to appear in the radiation area. This process enables the realization of different radiation modulation effects.

The intensity values can be obtained from an intensity detector (bucket detector) in each measurement, as shown in Eq. 1:

$$\boldsymbol{I}_i = \boldsymbol{P}_i * \boldsymbol{O} + \boldsymbol{\delta} \tag{1}$$

Where:

$\boldsymbol{P}_i$ represents $N^2$ modulation valves corresponding to the $i$-th sub-coding plate (a measured pattern) and it is transformed into a row vector.

**O** represents $N^2$ pixel values in an object and is transformed into a column vector.

**δ** represents the correlated noise.

Finally, an object image can be reconstructed with $P_i$ and $I_i$ by the traditional ghost imaging algorithms, as shown in Eq.2:

$$\tilde{O} = \langle I_i P_i \rangle - \langle I_i \rangle \langle P_i \rangle \tag{2}$$

Where:
$\tilde{O}$ represent the reconstructed result of an object.
$\langle . \rangle$ denotes the average of all measurements in CGI.

Later, with the development of imaging technology, orthogonal matching pursuit (OMP) and TVAL3 algorithm [27] based on compressive sensing theory have been applied successively and greatly improved the quality of reconstructed image.

**2.2 Rotational modulation method**

To achieve high-quality imaging of an object with N×N resolution, it is necessary to perform $N^2$ measurements. In the traditional modulation technique, if the approach of horizontal/vertical movement for switching sub-coding plates is employed, it would entail the fabrication of a large-scale coding plate with $N^4$ pixels. This is an exceedingly difficult and costly endeavor. Thus, there is a pressing need for a novel modulation technique that employs coding plates with a reduced number of pixels and delivers higher imaging quality. This is crucial for achieving high-resolution ray CGI.

The paper introduces a novel modulation technique called rotational modulation method. We achieve a reduction in the quantity of sub-coding plates by enhancing the utilization efficiency of each individual sub-coding plates, consequently leading to a decrease in the number of pixels on the coding plates. The specific method involves completing a measurement with a single sub-coding plate, followed by rotating the plate by an angle and conducting another intensity measurement. The rotation introduces changes in the spatial positions of sub-coding plate pixels, resulting in varying modulation effects (ray field distribution). As a result, we can achieve multiple intensity measurements by employing a strategy of rotating a sub-coding plate. This approach can greatly enhance the utilization efficiency of coding plate. Ultimately, only a small number of coding plates are required for the ray CGI. This approach significantly reduces the manufacturing quantity of sub-coding plates.

Certainly, the choice of coding plate for rotational measurements is a crucial consideration[20]. For rotational modulation method, our system matrix **P** is jointly determined by the pattern distribution (**B**) etched onto the sub-coding plate and the corresponding rotation angle. This relationship is expressed by the following equation:

$$F(\boldsymbol{B}_i, \theta) = \boldsymbol{P}_i \qquad (3)$$

Where:

$\boldsymbol{B}_i$ represents the sub-coding plate used in $i$-th measurement;

$\theta$ is the rotation angle during the i-th measurement.

In the rotational measurement method, what type of sub-coding plates ($\boldsymbol{B}_i$) should be chosen for conducting measurements? Currently, there is a lack of research addressing these aspects.

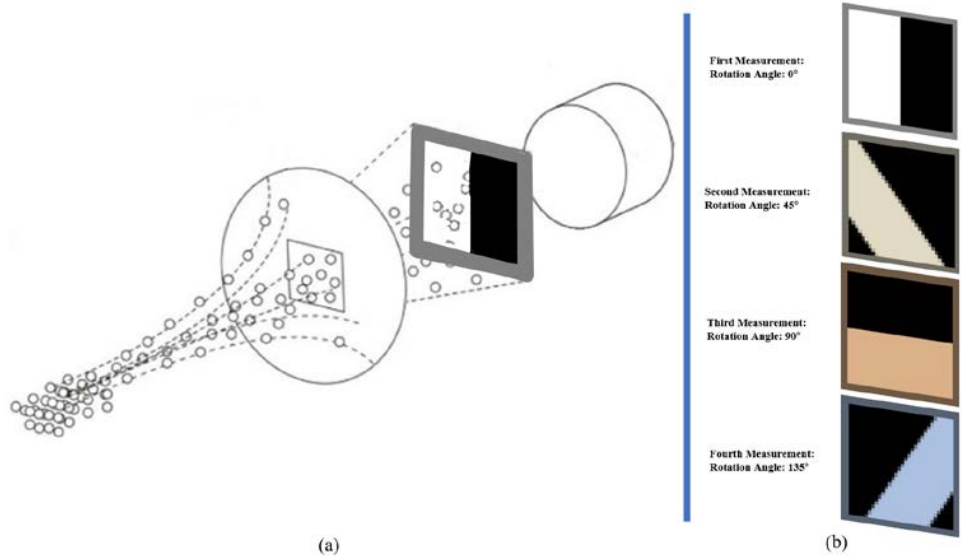

Fig.2 Schematic of rotational modulation method. (a) The measurement scenario is depicted when the rotation angle is set to 0 degrees. (b) The corresponding ray distribution is shown for different rotation angles of the sub-coding plate.

## 2.3 Single-stripe coding plate

We have discovered that the rotational measurement has been applied in computed tomography (CT). When performing cross-sectional imaging of an object, it is necessary to position a row of radiation sources at one end of the object and a row of detectors in a linear configuration at the opposite end.

Radiation beams generated from the radiation sources travel along a straight path through the object. During this journey, these beams undergo a continuous cumulative attenuation within the object. Ultimately these beams are detected by the detectors positioned on the opposite side. Following this, the system is subjected to a rotation by a specific angle and the measurement process continues. The measurement procedure is visually presented on Fig. 3 (a). If we approach the analysis of CT's measurement process from a computational imaging perspective, an intriguing revelation emerges: CT also represents a form of computational imaging.

Firstly, in CT, a measured intensity value encapsulates information about a portion of the object, specifically the material distribution within a certain columnar region of the cross-sectional slice. This phenomenon can be attributed to the fact that the attenuation of radiation within the object approximately follows the Beer-Lambert law. The law is described by the following equation:

$$I = I_0 \times e^{-\mu d} \tag{4}$$

Where:

$I_0$ represents initial ray intensity;

$I$ represents the ray intensity after passing through a pixel of the object;

$d$ represents length of a pixel;

$\mu$ represents the attenuation coefficient of the pixel.

In actual measurements, it is imperative to acknowledge that radiation particles emanating from a singular radiation source shall inevitably traverse numerous pixel regions within the targeted object, as shown in Fig. 3(b). Within this intricate journey, it becomes apparent that each pixel region possesses its own distinctive attenuation coefficient, denoted as $\mu_i$, with the subscript 'i' being representative of the corresponding pixel index. And the lengths of these pixels are also distinct, denoted as $d_i$.

According to the Eq. (4), the initial intensity of the ray beam is denoted as $I_0$. As this beam goes through each pixel, its intensity experiences a distinct form of attenuation, where exemplifications of such attenuated values can be represented as $I_1 = I_0 e^{-\mu_1 d_1}$、$I_2 = I_1 e^{-\mu_2 d_2}$、$I_3 = I_2 e^{-\mu_3 d_3}$, and so forth. Consequently, the final intensity value measured by the detector can be expressed in Eq. (5).

$$I_{det} = I_0 e^{-(\mu_1 d_1 + \mu_2 d_2 + \mu_3 d_3 + \cdots + \mu_n d_n)} \tag{5}$$

Where:

$I_{\text{det}}$ represents the measured value obtained by a single detector in CT;

$I_0$ represents the initial ray intensity.

$\mu_1, \mu_2 \cdots \mu_n$ represents the attenuation coefficient of the ray passing through the pixel.

$\mu_{n+1}, \mu_{n+2} \cdots \mu_{N \times N}$ represents the attenuation coefficient that has not been passed through the pixels by the ray beam.

The attenuation of ray beam in the object follows an exponential decay. A logarithmic transformation can be employed to transform measured value to CT projection value $I^{CT}$, as shown in the following equation:

$$I^{CT} = \log \frac{I_0}{I_{det}} = \mu_1 d_1 + \mu_2 d_2 + \mu_3 d_3 + .... + \mu_n d_n \qquad (6)$$

Secondly, in the realm of computational imaging, its each measured intensity values also possesses information about a portion of the object. Therefore, we can attempt to view a measurement value within CT as an equivalent to a measurement in computational imaging.

Although CT doesn't utilize sub-coding plate to directly modulate the radiation field, the arrangement where each radiation source corresponds to a detector ensures that the radiation passes through only a single columnar region of the object. This arrangement is equivalent to the placement of a sub-coding plate with a single column of apertures, as shown in Fig. 3(c). Indubitably, this approach artfully accomplishes the indirect modulation of the radiation field.

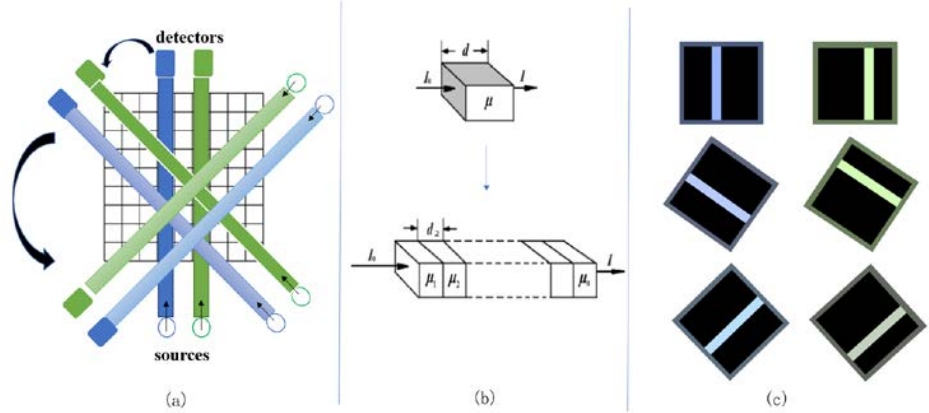

Fig.3 Schematic diagram of CT measurements. (a) CT measurement process. The object is divided into a 9×9 pixels. The radiation sources and detectors are positioned along a line. The blue and green regions represent paths that ray beams emitted by different sources go through. After a measurement is completed, both detectors and sources are synchronously rotated counterclockwise by an angle to make the next measurement; (b) attenuation process of radiation; (c) sub-coding plates with a single column of apertures. Two distinct-color sub-coding plates are employed, corresponding to the two paths in (a). Different color depths represent measurements at different angles.

Therefore, the CT measurements can be reformulated into an expression corresponding to computational imaging. Here, $\mu_j$ encompasses the information of the $j$-th pixel in the object, and thus $\mathbf{O}(j) = \mu_j$. The lengths ($d_1 \cdots d_n$) of the rays passing through various pixels during the $i$-th measurement as modulation effects of the rays are filled into the system matrix ($P_i^{CT}$). So, the measurement value ($I_i^{CT}$) for the *i*-th measurement in CT can be expressed as follows:

$$I_i^{CT} = P_i^{CT} \times \mathbf{O} + \delta \qquad (7)$$

Thirdly, due to the integrity of the CT measurement data, the object can be fully recovered with all measured values of CT. Therefore, when employing the rotational measurement scheme proposed in this paper for CGI, high-quality imaging is possible if the chosen coding plate can achieve modulation effects that are similar to $P^{CT}$. On the other hand, in CT, the measurement value is solely dependent on the interaction between a column region of the object and the rays. Therefore, the corresponding modulation effect is achievable by utilizing some single-column aperture sub-coding plates and rotating those at specific angles.

The following diagram will display the shape of this type sub-coding plates, which we refer as "single-stripe coding plate".

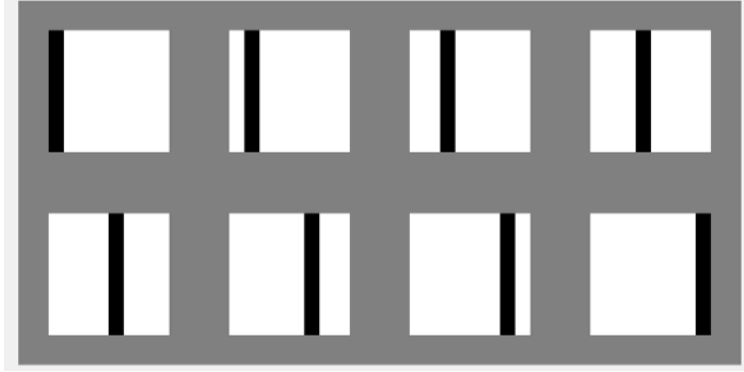

Fig.4 Single-stripe coding plate. It is a set of single-stripe coding plates required for an 8×8 resolution object. This assembly comprises 8 distinct sub-coding plates. Each sub-coding plate has only one hollowed-out column area (depicted as a black area) that allows ray to pass through directly. During rotational modulation method, it is necessary to sequentially employ these 8 sub-coding plates. Each sub-coding plate needs to be individually rotated at eight different angles to acquire eight measurement values. After completing these measurements, the next sub-coding plate is replaced to continue the measurement process.

## 2.4 Calculation of system matrix

In the rotational modulation method, the system matrix is no longer a traditional binary matrix, meaning $B_i \neq P_i$. It becomes crucial that precisely calculating the system matrix for a single-stripe coding plate with rotational modulation method. At some angle, after the ray source is modulated by a striped sub-coding plate, only one column of ray beam can pass through, while other rays passing through other areas of the sub-coding plate will be blocked by its material, as shown in Fig. 5. Therefore, we only need to calculate the corresponding area of this column ray beam in each pixel of the object. To be noticed, the pixels passed by the beam at different angles are different. How to accurately calculate corresponding area in each pixel of the object (modulation matrix) to represent the modulation effect of the radiation field under rotation measurement is a very important work.

In the calculation of system matrix, all the pixels of the object are divided into three cases: (1) there is no intersection between the pixel and the beam. (2) the whole pixel is inside the beam. (3) the pixel intersects with the beam boundary, that is, a part of the pixel is inside the beam. Thus, it is necessary to obtain the case of each pixel.

An area model based on beam boundaries is established. The pixels between upper and lower boundary are the second type of pixels, and their intersecting areas are 1. The pixels which are above the upper boundary or below the lower is the first case, and their intersecting areas are 0. The areas in the second case are the most difficult to calculate. Next, we will introduce how to calculate their areas in detail.

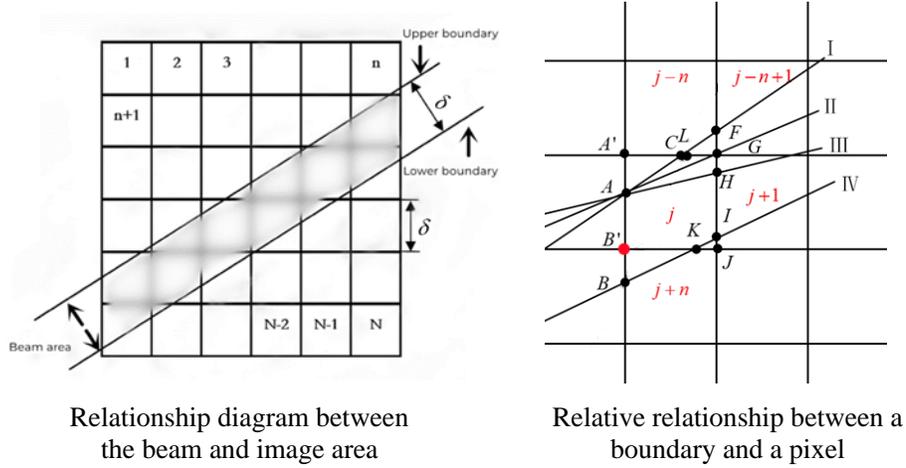

Relationship diagram between the beam and image area

Relative relationship between a boundary and a pixel

Fig.5 Beam model. In the left, the image area is divided into N=n x n pixels, and the side length of each pixel is $\delta$. The shaded area is the beam which represents a distribution of residual ray field modulated by a striped sub-coding plate. In the right, The five intersection kinds (I, II···VI) of a boundary and No. J pixel (B'A'GJ) are shown. The red font indicates the next possible pixel which the beam will pass through.

Firstly, in the rotation modulation of the stripe sub-coding plate, there are many kinds of intersections between the beam and the image pixels. And the calculating forms of the intersection areas are different in different kinds.

We assume that the boundary equation is $y = mx + b$, and two intercepts $(d_1, d_2)$ between a boundary and pixel whose lower-left corner coordinates are $(x_{B'}, y_{B'})$ can be calculated by the following formula:

$$d_1 = mx_{B'} + b - y_{B'} \tag{8}$$

$$d_2 = m(x_{B'} + \delta) + b - y_{B'} \tag{9}$$

Then according to the two intercepts, the intersection of the boundary and the pixel can be divided into four kinds(I, II…VI). The specific kinds and classification standards are shown in the Fig.6 and table 1, respectively.

Table 1 Classification standards of four kinds

| Kinds | $d_1$ | $d_2$ | Next pixel number |
|---|---|---|---|
| I | $d_1 \geq 0$ | $d_2 > \delta$ | J-N |
| II | $d_1 \geq 0$ | $d_2 = \delta$ | J-N-1 |
| III | $d_1 < 0$ | $d_2 \leq \delta$ | J+1 |
| VI | $d_1 < 0$ | $d_2 \leq \delta$ | J+1 |

Secondly, in the five kinds, if the beam boundary is the upper boundary, the intersection area between the beam and a pixel are as follows:

Case I:

$$S_{ACGJB'} = \delta^2 - \frac{(\delta - d_1)^2}{2m} \quad (10)$$

Case II:

$$S_{AHJB'} = \delta^2 - \delta \frac{\delta - d_1}{2} \quad (11)$$

Case III:

$$S_{B'AHJ} = \delta^2 - \delta \frac{2\delta - d_1 - d_2}{2} \quad (12)$$

Case VI:

$$S_{KIJ} = \frac{d_2^2}{2m} \quad (13)$$

If the beam boundary is the lower boundary, the intersection area between the beam and a pixel are as follows:

Case I:

$$S_{AA'C} = \frac{(\delta - d_1)^2}{2m} \quad (14)$$

Case II:

$$S_{AA'G} = \delta \frac{\delta - d_1}{2} \tag{15}$$

Case III:

$$S_{AA'GH} = \delta \frac{2\delta - d_1 - d_2}{2} \tag{16}$$

Case VI:

$$S_{B'A'GIK} = \delta^2 - \frac{d_2^2}{2m} \tag{17}$$

Thirdly, based on the interaction kinds between the beam and a pixel, we obtain all the intersection areas between the entire beam and all image pixels. Due to axis symmetry, we only need to consider the case of $0<m\leqslant 1$. The calculation process begins at the first pixel where the beam boundary intersects the object. The intersecting area can be calculated according to $d_1$ and $d_2$. Then according to Table 1, the next pixel can be determined and proceed to the next round of calculation. Until the beam leaves the image. The intersecting area values of all pixels are used to form a modulation matrix ($P_i^{CT}$) corresponding to a measurement. Finally, all the modulation matrices can be obtained and form a system matrix ($\mathbf{P}^{CT}$) of the CGI.

## 3. Simulation and experimental setup

In the paper, the simulation and experiment are employed to validate the feasibility of rotational modulation method with the single-stripe coding plate. In the simulation phase, we conducted experiments using two types of data, obtained from numerical and Monte Carlo simulations. In the numerical simulation, an object shaped as the characters 'GI' with 32×32 pixels is employed and the intensity data under full-sampling and under-sampling is obtained by matrix multiplication.

However, the process of ray CGI involves intricate interactions between particles and matter, which cannot be accurately depicted by traditional numerical simulations. Geant4 is a powerful Monte Carlo simulation toolkit. It can be used to simulate the physical process of particle transport in matter. [28] It is widely used in high-energy physics, nuclear physics, medical physics, and radiation protection. The data from Geant4 simulations were more realistic than from numerical simulation. Therefore, a ray CGI platform using Geant4 was established and data very similar to the real experimental conditions were acquired. This Monte Carlo simulation aims to further verify the feasibility and practical imaging effectiveness of our proposed scheme.

The Monte Carlo simulation diagram in our scheme is shown in Fig. 5(a). The bucket detector, measured object, sub-coding plate and radioactive source were placed on the same horizontal plane. The particles were emitted from radioactive sources, then pass through a sub-coding plate and object, and finally were received by the bucket detector. The object is a middle-perforated metal block, the side length of this metal block is 32 cm, and the aperture radius is 4 cm (4 pixels), as shown in Fig.6 (b).

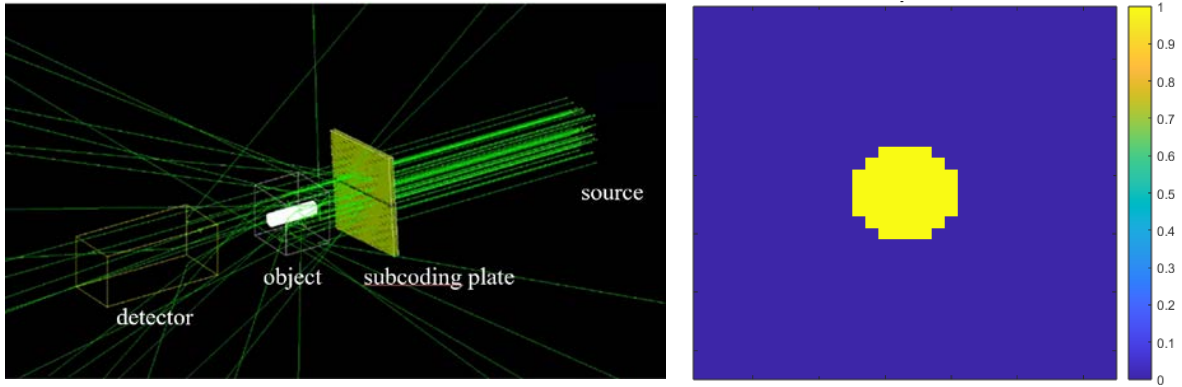

(a) Monte Carlo simulation diagram                    (b) Original image

Fig.6 Monte Carlo simulation diagram and original image

In the practical experiment phase, we embarked upon the creation of an experimental platform of α ray CGI. The specific configurations are shown in the Fig. 7.

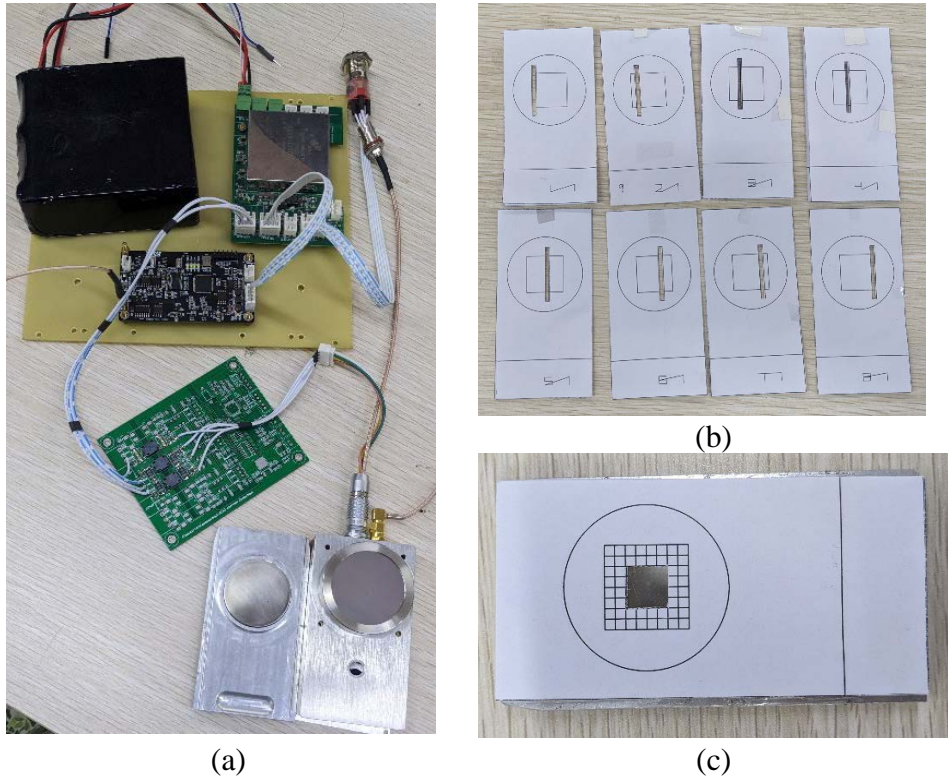

Fig.7 Experimental platform of α ray CGI. (a) Detection equipment. It encompasses three fundamental components: the intensity measurement system, the α ray areal source. The intensity detector is consisted of a PIPS (Passivated Implanted Planar Silicon) probe, front-end electronics, and channel amplitude analysis section. This detector can obtain an intensity value from ray energy spectrum. But the detector lacks spatial resolution capability. The radiation source is an α source ($Pu^{238}$) with an activity of $1.96\times10^3$ Bq and a diameter of 4 cm. (b) Single-stripe sub-coding plates. These sub-plates are crafted by papers. (c) A partially occluded radiative source. It is formed by placing a paper sheet with a central aperture over the α ray areal source. The circular region on the paper represents the area that the detector can detect, and the square region represents the region for imaging.

## 4. Results and discussions
### 4.1 Numerical simulation

Numerical simulations were primarily conducted to validate the feasibility of the proposed rotational modulation method in this paper. And reconstructed image quality of the rotational modulation method was compared with that of a Hadamard scheme. Under the full-sampling condition, a mere 32 individual single-stripe sub-coding plates was necessitated in the rotational modulation scheme. Each sub-coding plate was continuously rotated at 32 different angles
14

for a total of 32 measurements. In the Hadamard scheme, 1024 sub-coding plates were utilized. Under the under-sampling condition, a sampling rate of 20% was set. In accordance with our proposed scheme, each individual sub-coding plates could only perform 7 measurements. Meanwhile, sub-coding plates based on the Russian Doll (RD) sampling sequence were employed in the Hadamard scheme. Ultimately, we used the TVAL3 algorithm to get reconstructed image of all schemes, which were shown in Fig. 8. In addition, the objective evaluation index - peak signal to noise ratio (PSNR) was employed to quantitatively evaluate the quality of all reconstructed images. And its formula was as follows:

$$PSNR = 10 \times \lg(\frac{MAX_O^2 \times N^2}{\sum_{i=1}^{N}\sum_{j=1}^{N}(x(i,j) - \hat{x}(i,j))^2}) \qquad (18)$$

Where:
$x(i,j)$ represents a gray value of an original image at $(i,j)$ pixel;
$\hat{x}(i,j)$ represents a gray value of a reconstructed image at $(i,j)$ pixel.

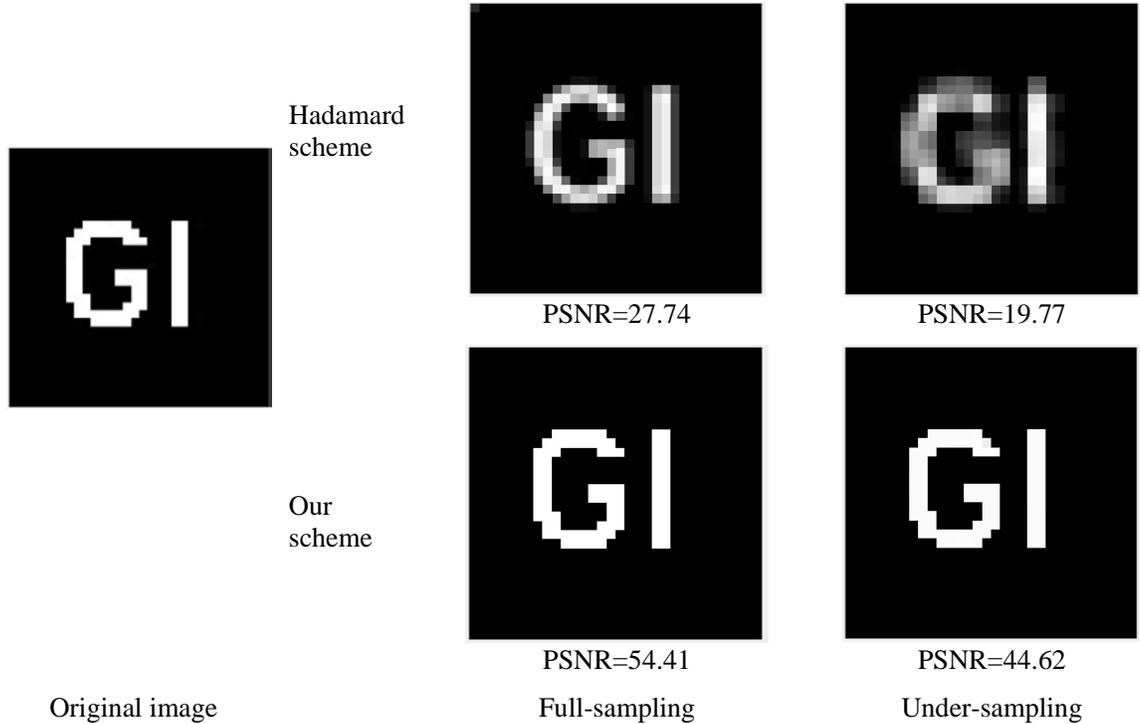

Fig. 8 Results of numerical simulation

Firstly, under the full-sampling, our scheme subjectively yielded superior imaging quality compared to the Hamdard scheme. In the reconstructed images obtained from our scheme, the boundary details of the "GUI" were clearly shown and there was a little noise. The imaging quality was exceptionally high, and its PSNR was up to 54.41. Conversely, in the Hadamard scheme, there existed a certain degree of blurriness in both the main body and the boundaries of the "GUI", leading to a reduction in imaging quality. Its PSNR was only 27.74. Consequently, for the binary objects, the reconstructed results from our scheme were superior to those from the Hadamard scheme in both subjective and objective evaluations.

Additionally, in the Hadamard scheme, comparing to results of full-sampling, the edges of the reconstructed object were noticeably blurred and the image quality was noticeably degraded. The PSNR value decreased to 19.77, representing a significant decrease of 28%. In contrast, our scheme could accurately recover objects under under-sampling conditions. And its PSNR was still up to 44.62. So those conclusions in full-sampling were consistently validated in the under-sampling condition. One of the reasons for these improvements is that the distribution of modulated ray fields $P_i$ in our scheme follows a grayscale matrix. This characteristic allows the measured values to encapsulate more information of object, consequently leading to superior image recovery results.

In summary, the rotational modulation method combined with a single-stripe coding plate not only achieves high-quality imaging under full sampling conditions, but also maintains its effectiveness under under-sampling conditions.

### 4.1 Monte Carlo Simulation

To further verify the feasibility of our scheme and the conclusions obtained in numerical simulation, we carried out the Monte Carlo simulation experiments. The results were shown as follow:

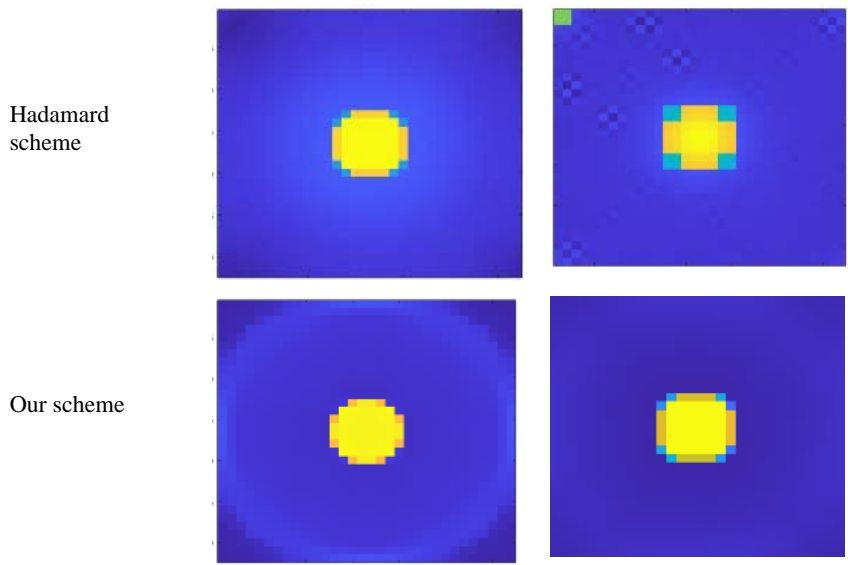

Full-sampling　　　　　　Under-sampling

Fig.9 Result of Monte Carlo simulation

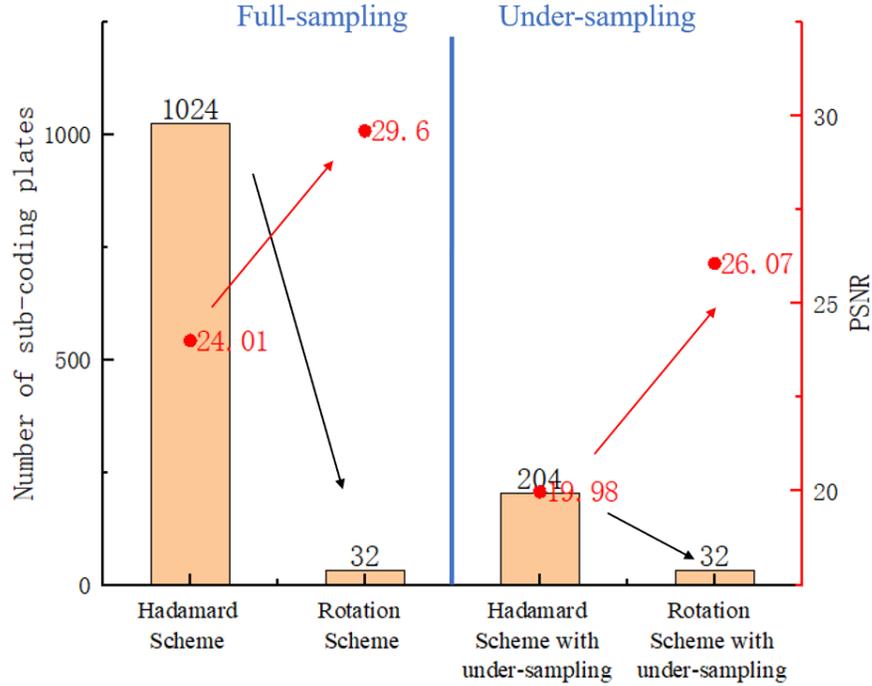

Fig.10 Quantitative evaluation results of reconstructed images

As shown in Fig. 9, our proposed scheme consistently achieves high-quality imaging results when employed in full-sampling and under-sampling. For example, when comparing our scheme with the Hadamard scheme in under-sampling, our scheme accurately reconstructed the edge of the hole.

Furthermore, the comparison in Fig. 10 shown that our scheme outperforms the Hadamard scheme in terms of image quality and the number of sub-coding plates required. Under full sampling, our scheme achieved a higher PSNR of 29.6 with only 32 sub-coding plates, while the Hadamard scheme required 1024 plates to achieve a PSNR of 24.01. Similarly, under under-sampling, our scheme again outperforms the Hadamard scheme. Our scheme achieved high-quality imaging with only 32 sub-coding plates, while the Hadamard scheme still required 204 plates.

Overall, our scheme enables high-quality imaging with a minimal number of sub-coding plates, making it more efficient and practical compared to the Hadamard scheme.

## 4.3 Experiment

In this study, we conducted experiments using our proposed scheme on the experimental platform mentioned in the previous section. The imaging was performed at a resolution of 8×8. The imaging object is a partially shielded radiative source, as illustrated in Fig.7(c). The result was shown in follow.

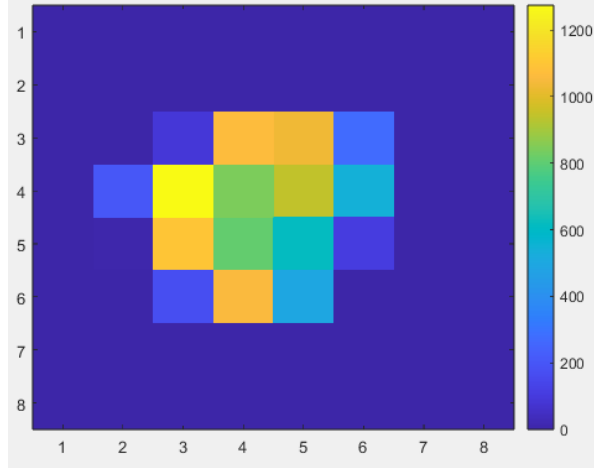

Fig.11 Reconstructed imaging

In the Fig. 11, our scheme successfully reconstructs the intensity distribution of the unobstructed region. This region appears as a 4x4 square, which is similar with the shape of paper sheet on source. However, it is worth noting that the imaging quality diminishes at the edges of the region. For example, at position (4,2), the ideal image should display no intensity, indicating 100% blocking of the rays. The actual image of this position shows a slight intensity. This phenomenon can be attributed to the uneven distribution of source intensity, resulting in higher ray intensity in this position. The rays of this position may not be completely obstructed by the paper sheet.

In summary, our scheme, utilizing only a little sub-coding plates and a single-pixel ray intensity detector, approximately reconstructs the object. This advancement contributes to the cost-effectiveness of ray imaging technology and offers a novel approach for downsizing coding plates.

## 5. Conclusion

In this paper, we proposed a novel form of rotation measurement, which can realize ray ghost imaging with few sub-coding plates. A beam model was constructed to calculate the system matrix. Then the simulation and actual experiments are carried out. Those experimental results shown that our scheme can achieve a clear image no matter under full-sampling or under-sampling. And in the quantitative analysis, our scheme is also better than Hadamard scheme. More importantly, the number of sub-coding plates required for our scheme is much smaller than the Hadamard scheme. Overall, our scheme offers significant advantages in imaging quality, can greatly reduce the requirement of imaging system on the number of sub-coding plates and the difficulty of manufacturing the coding plate. Thus, this work contributes to the realization of ray CGI.

**Acknowledgments**

The authors would like to acknowledge the support of the Youth Science Foundation of Sichuan Province (No.2022NSFSC1230 and No.2022NSFSC1231), The General project of national Natural Science Foundation of China (No.12075039) and The Key project of the National Natural Science Foundation of China (No. U19A2086).

**Disclosures**

The authors declare no conflicts of interest.

applications to the single pixel camera and compressive sensing, Rice University, 2010.